\documentclass[aip,apl,reprint,noshowpacs,noshowkeys,amssymb,amsmath,amsfonts,bibnotes]{revtex4-2}

\usepackage{graphicx}
\usepackage{theorem}
\usepackage{dcolumn}
\usepackage{longtable}
\usepackage{bm}
\usepackage{accents}
\usepackage{color}
\usepackage[dvipsnames]{xcolor}
\usepackage{makecell}
\usepackage{lineno}

\renewcommand{~}{\,}

\makeatletter
\renewcommand{\figurename}{Fig.}
\renewcommand{\tablename}{Tab.}
\renewcommand{\fnum@figure}[1]{\textbf{\figurename~\thefigure.} }
\renewcommand{\fnum@table}[1]{\textbf{\tablename~\thetable.} }
\makeatother

\begin{document}

\title{Coherent manipulation of interacting electron qubits on solid neon}

\author{Xinhao Li}\email[Email: ]{xinhaoli@fas.harvard.edu}
\affiliation{Department of Physics, Harvard University, Cambridge, Massachusetts 02138, USA\looseness=-1}
\affiliation{Center for Nanoscale Materials, Argonne National Laboratory, Lemont, Illinois 60439, USA\looseness=-1}

\author{Yizhong Huang}
\affiliation{Center for Nanoscale Materials, Argonne National Laboratory, Lemont, Illinois 60439, USA\looseness=-1}
\affiliation{Pritzker School of Molecular Engineering, University of Chicago, Chicago, Illinois 60637, USA\looseness=-1}

\author{Xu Han}
\affiliation{Center for Nanoscale Materials, Argonne National Laboratory, Lemont, Illinois 60439, USA\looseness=-1}
\affiliation{Pritzker School of Molecular Engineering, University of Chicago, Chicago, Illinois 60637, USA\looseness=-1}

\author{Xianjing Zhou}\email[Email: ]{xianjing.zhou@fsu.edu}
\affiliation{National High Magnetic Field Laboratory, Tallahassee, Florida 32310, USA}
\affiliation{Department of Mechanical Engineering, FAMU-FSU College of Engineering, Florida State University, Tallahassee, Florida 32310, USA}

\author{Amir Yacoby}
\affiliation{Department of Physics, Harvard University, Cambridge, Massachusetts 02138, USA\looseness=-1}

\author{Dafei Jin}\email[Email: ]{dfjin@nd.edu}
\affiliation{Department of Physics and Astronomy, University of Notre Dame, Notre Dame, Indiana 46556, USA\looseness=-1}

\date{\today}

\begin{abstract}

\end{abstract}

\maketitle
\pretolerance=9000 

\textbf{Electrons trapped on solid neon surfaces serve as low-noise charge qubits with long coherence times and high operational fidelities. Such charge qubits offer full electrical control and compact device footprints, convenient for scaling up with quantum circuits. Realizing two-qubit gates on this platform is a critical step towards practical quantum information processing. In this work, we report the first experimental demonstration of coherent manipulation of multiple interacting electron-on-solid-neon (eNe) charge qubits. By exploiting the electrons naturally confined in close proximity by the surface structures of solid neon, we have achieved a direct qubit-qubit coupling strength of up to 62.5\,MHz, as well as implemented cross-resonance (CR) and bSWAP two-qubit gates using global microwave drives. The natural electron confinement by solid neon mitigates the high-density-wiring challenge, simplifies the multi-qubit control, and establishes a unique path to scale up the eNe qubit platform.}\\

Scalable quantum processors require the constituent qubits to have long coherence, fast and accurate gates, and high-connectivity integration. To that extent, electron-based qubits have unique advantages for large-volume solid-state system scaling, thanks to electrons’ small footprints. For example, spin qubits based on electrons in gate-defined quantum dots or implanted dopants exhibit very long coherence, and form multi-qubit arrays~\cite{burkard2023semiconductor, Petta2005coherent, borsoi2024shared,vandersypen2017interfacing, zwerver2022qubits, hu2025single}. However, this scaling requires complex gate stacks for electron confinement and spin detection at the individual qubit scale. Meanwhile, the necessity for global or local magnetic field gradients to address electron spin adds another layer of hardware complexity for integration. Those overheads impede the volume growth of electron-based qubit arrays~\cite{vandersypen2017interfacing, borsoi2024shared}.

Electron-on-solid-neon (eNe) offers an emerging solid-state qubit platform that combines excellent coherence and minimal device overhead~\cite{zhou2022single, zhou2024electron, guo2024quantum, jennings2024quantum}. In this system, single electrons sit in a near-vacuum environment a few nanometers above the neon surface, where the noble-gas substrate provides superior noise isolation~\cite{li2025noise}. Qubits based on the charge (motional) states of eNe have shown exceedingly long coherence~\cite{zhou2024electron}. The inherent electrical addressability of charge states allows for straightforward and high-fidelity single-qubit gates with microwave photons in a circuit quantum electrodynamics (cQED) architecture~\cite{schuster2010proposal, blais2021circuit}, significantly reducing the hardware complexity compared to electron spin qubits.

To further advance the eNe platform towards a functional quantum information architecture, it is essential to construct systems with multiple interacting qubits and realize their coherent manipulation~\cite{platzman1999quantum, jennings2024quantum}. Previously, we observed two eNe charge qubits co-trapped on the same device and simultaneously coupled to a common superconducting resonator~\cite{zhou2024electron}. However, the through-resonator inter-qubit coupling was not strong enough for coherent real-time two-qubit gate operations. Nonetheless, recent theoretical and experimental studies suggest that solid-neon surface morphology offers a natural mechanism to trap multiple electrons at close distances~\cite{kanai2024single, li2025noise, zheng2025surface}. It shows the promise of creating strong interactions between adjacent qubits maintained solely by the neon surface. This can greatly simplify the originally complex electrodes in typical electron-based qubit systems~\cite{vandersypen2017interfacing} and reduce gate-introduced electrical noise.

\begin{figure*}[ht]
\centerline{\includegraphics[scale=0.5]{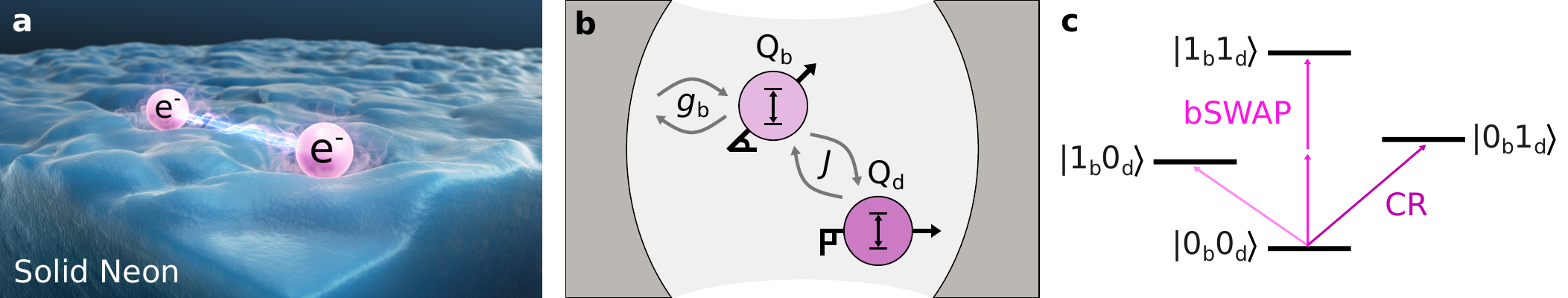}}
\caption{\textbf{Schematic of multi-qubit coupling on solid neon.} \textbf{a,} Solid neon creates multi-qubit host spots and supports inter-qubit coupling at a short distance. \textbf{b,} Schematic of the interactions between a two-qubit system on solid neon and a superconducting resonator, where a bright qubit $\mathrm{Q_b}$ and a dark qubit $\mathrm{Q_d}$ are coupled with strength $J$, and $\mathrm{Q_b}$ is coupled to the resonator with strength $g_\mathrm{b}$. Due to the orthogonal dipole alignment to the field, as marked by the black arrow, $\mathrm{Q_d}$ is invisible to the resonator. \textbf{c,} Energy-level schematic of the two-qubit system in \textbf{b}, illustrating the possible transitions. Both the cross-resonance (CR) and the bSWAP type two-qubit operations are shown, where the $\mathrm{Q_d}$ are driven by microwave gates mediated through the $\mathrm{Q_b}$.}
\label{Fig:two_qubit_picture}
\end{figure*}

In this work, we explore this novel potential offered by solid neon and achieve the first coherent manipulation of multiple interacting electron qubits on this platform. We utilize the neon surface structure as a natural mechanism to pack several electrons at close distances to establish strong charge-charge interactions~\cite{beysengulov2024coulomb, jennings2024quantum}. Further, we prove that the natural confinement of electrons by solid neon offers a unique engineering resource to qubit-resonator and inter-qubit coupling, enabling multi-qubit operation with simplified wiring. In a two-qubit system, spectroscopic measurements show an inter-qubit coupling strength of $J/2\pi$ = 3.35\,MHz, surpassing the qubits' decoherence rates. By driving one of the two qubits, which is the only one strongly coupled with the superconducting resonator, we demonstrate the cross-resonance and bSWAP types of all-microwave two-qubit operations~\cite{krantz2019quantum}, with just a single direct-current (DC) qubit bias. The experimental results match well with the theoretical simulations based on the measured parameters. Furthermore, we observe even stronger coupling in a three-qubit system, with a maximum qubit interaction strength of $J/2\pi$ = 62.5\,MHz. Together with previously achieved long coherence and noise isolation, this result on coherent multi-qubit manipulation completes the eNe platform with crucial components to overcome key scaling challenges faced by electron-based qubits.
\\

\noindent\textbf{Multi-qubits on solid neon}

The solid neon film can support the trapping of multiple electrons within a compact area without introducing complex electrical gate stacks at the individual qubit scale. Experimental and theoretical investigations show that surface structures at the 10-nm scale facilitate electron trapping on solid neon~\cite{zheng2025surface, kanai2024single}. Those surface structures can originate from the transfer of the underneath substrate roughness or the solidification process during the neon layer growth~\cite{leiderer2025surface}. As illustrated in Fig.~\ref{Fig:two_qubit_picture}a, the solid neon film naturally supports closely spaced electron traps and enables local qubit coupling through strong charge-charge interactions. 

The trapping of electrons on solid neon films also creates the possibilities for interaction engineering with simplified circuit wiring. In the cQED architecture, electrons can be driven and read out through microwave photons in the resonator. The electron-photon coupling strength is determined by the magnitude of the qubit's dipole moment and its alignment with the local microwave (MW) electric field. The neon surface morphology can lead to distinct electron-photon couplings~\cite{koolstra2019coupling, schuster2010proposal}. As depicted in Fig.~\ref{Fig:two_qubit_picture}b, inside a closely spaced two-qubit coupled system, one of the qubits with a dipole orthogonal to the MW field of a superconducting resonator exhibits a near-zero qubit-resonator coupling strength and negligible resonator dispersive shift.
We refer to this qubit as the ``dark" qubit $\mathrm{Q_d}$.
The other qubit, which has a dipole significantly aligned to the MW field, is coupled to the resonator with a strength of $g_\mathrm{b}$. This qubit is referred to as the ``bright" qubit $\mathrm{Q_b}$ and its quantum states can be read out dispersively. The charge-charge interaction between the two qubits results in the inter-qubit coupling strength $J$.

\begin{figure*}[ht]
\centerline{\includegraphics[scale=0.19]{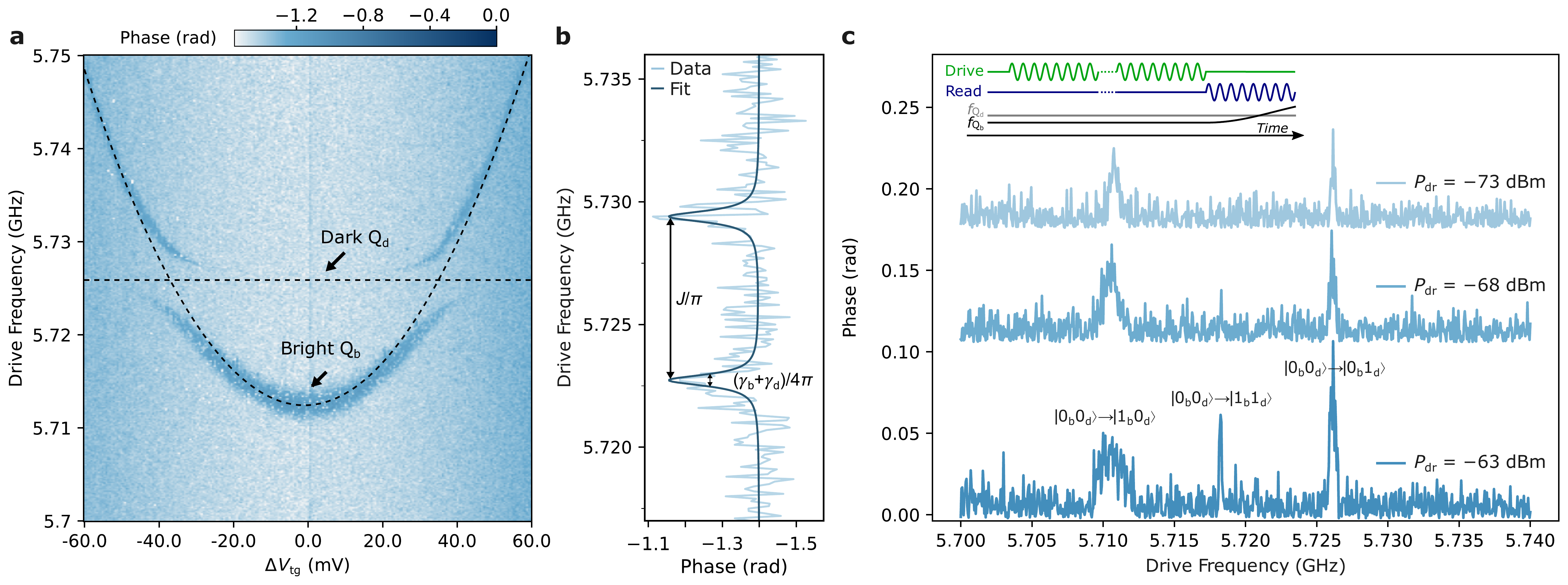}}
\caption{\textbf{Spectroscopic characterization of coherent two-qubit coupling on solid neon.} \textbf{a,} Continuous-wave (CW) two-tone qubit spectroscopy with avoided crossings on the resonator's phase response. The dashed lines indicate the bare transition frequencies of the dark qubit $\mathrm{Q_d}$ and the bright qubit $\mathrm{Q_b}$. \textbf{b,} The line cut at the avoided crossing shows a splitting corresponding to a charge-charge interaction strength of $J/2\pi=3.35$\,MHz, which exceeds the hybrid linewidth $(\gamma_\textrm{b}+\gamma_\textrm{d})/4\pi=0.56$\,MHz. \textbf{c,} Qubits spectroscopy measured with high-power pulsed readout, in which $\mathrm{Q_b}$ served as the mediator between $\mathrm{Q_d}$ and the resonator. The inset shows the pulse sequence: a long square pulse (10\,\textmu s) with variable frequency $f_\mathrm{dr}$ and power $P_\mathrm{dr}$ drive the system to a specific state, followed by a short square probe pulse (0.7\,µs) at the resonator frequency $f_\mathrm{r}$ with approximately $-120$\,dBm power reaching the resonator. The\ readout pulse induces the frequency crossing of the two qubits and the swap of their populations, revealing the state information of the dark qubit $\mathrm{Q_d}$. Drive pulse power $P_{\text{dr}}$ reaching the resonator's input coupler increases from $-73$\,dBm to $-63$\,dBm, activating transitions corresponding to the CR ($|0_\mathrm{b}0_\mathrm{d}\rangle \rightarrow |0_\mathrm{b}1_\mathrm{d}\rangle$) and bSWAP ($|0_\mathrm{b}0_\mathrm{d}\rangle \rightarrow |1_\mathrm{b}1_\mathrm{d}\rangle$) two-qubit operations. The measurements were taken when the system was biased at $\Delta V_{\rm tg}=~$0\,V. The phase curves are off-set for visualization.}
\label{Fig:two_qubit_spectrum}
\end{figure*}

The structurally induced dipole orientations distribution enables coherent multi-qubit operations to be driven with only global control, reducing the complexity of the local gate stacks. The two-qubit system described above supports all-microwave-driven two-qubit operations, including cross-resonance (CR) and bSWAP gates, as presented in Fig.~\ref{Fig:two_qubit_picture}c. All-microwave gates are widely used for fixed-frequency qubits thanks to their easy implementation, which also preserves qubit coherence by avoiding electrostatic and flux bias~\cite{krantz2019quantum}. For the CR gate, $\mathrm{Q_d}$ is excited by driving $\mathrm{Q_b}$ at the frequency of $\mathrm{Q_d}$, generating the transition $|0_\mathrm{b}0_\mathrm{d}\rangle \rightarrow |0_\mathrm{b}1_\mathrm{d}\rangle$ (refs.~\cite{rigetti2010fully,magesan2020effective,chow2011simple}). For the bSWAP gate, $\mathrm{Q_b}$ and $\mathrm{Q_d}$ are driven monochromatically through a two-photon process near the frequency middle point between the two, generating the transition $|0_\mathrm{b}0_\mathrm{d}\rangle \rightarrow |1_\mathrm{b}1_\mathrm{d}\rangle$ (refs.~\cite{poletto2012entanglement,roth2017analysis}).\\

\noindent\textbf{Two-qubit spectroscopy}

We have experimentally achieved strong charge-charge coupling between eNe qubits, assisted by the neon surface, as proposed. See Methods for the details of the superconducting microwave resonator, neon growth and electrons deposition. A two-qubit system was probed by the continuous wave (CW) two-tone qubit spectroscopy while tuning one electrode gate ($\Delta V_{\rm tg}$, see Methods). Two clear avoided crossings which indicate the qubit-qubit strong coupling appear in the spectrum in Fig.~\ref{Fig:two_qubit_spectrum}a. One of the qubits is visible in the CW spectroscopy and referred to as the ``bright" qubit $\mathrm{Q_b}$, while the other non-detectable one is denoted as the ``dark" qubit $\mathrm{Q_d}$. The magnitude of the splitting represents an inter-qubit coupling strength of $J/2\pi$ = 3.35\,MHz, exceeding the linewidth of the hybrid states $(\gamma_\textrm{b}+\gamma_\textrm{d})/4\pi$ = 0.56\,MHz, as shown in Fig.~\ref{Fig:two_qubit_spectrum}b.

Individually, $\mathrm{Q_b}$ has a hyperbolic frequency dependence on $\Delta V_{\rm tg}$, with a charge sweet spot (SS) near 5.71\,GHz, which is about 40\,MHz above the resonator frequency. 
We estimate the coupling strength between $\mathrm{Q_b}$ and resonator to be $g_\mathrm{b}/2\pi$ = 3.76\,MHz, based on the dispersive pushing at the qubit's SS (Extended Data Fig.~1).

\begin{figure*}[htb]
	\centerline{\includegraphics[scale=0.36]{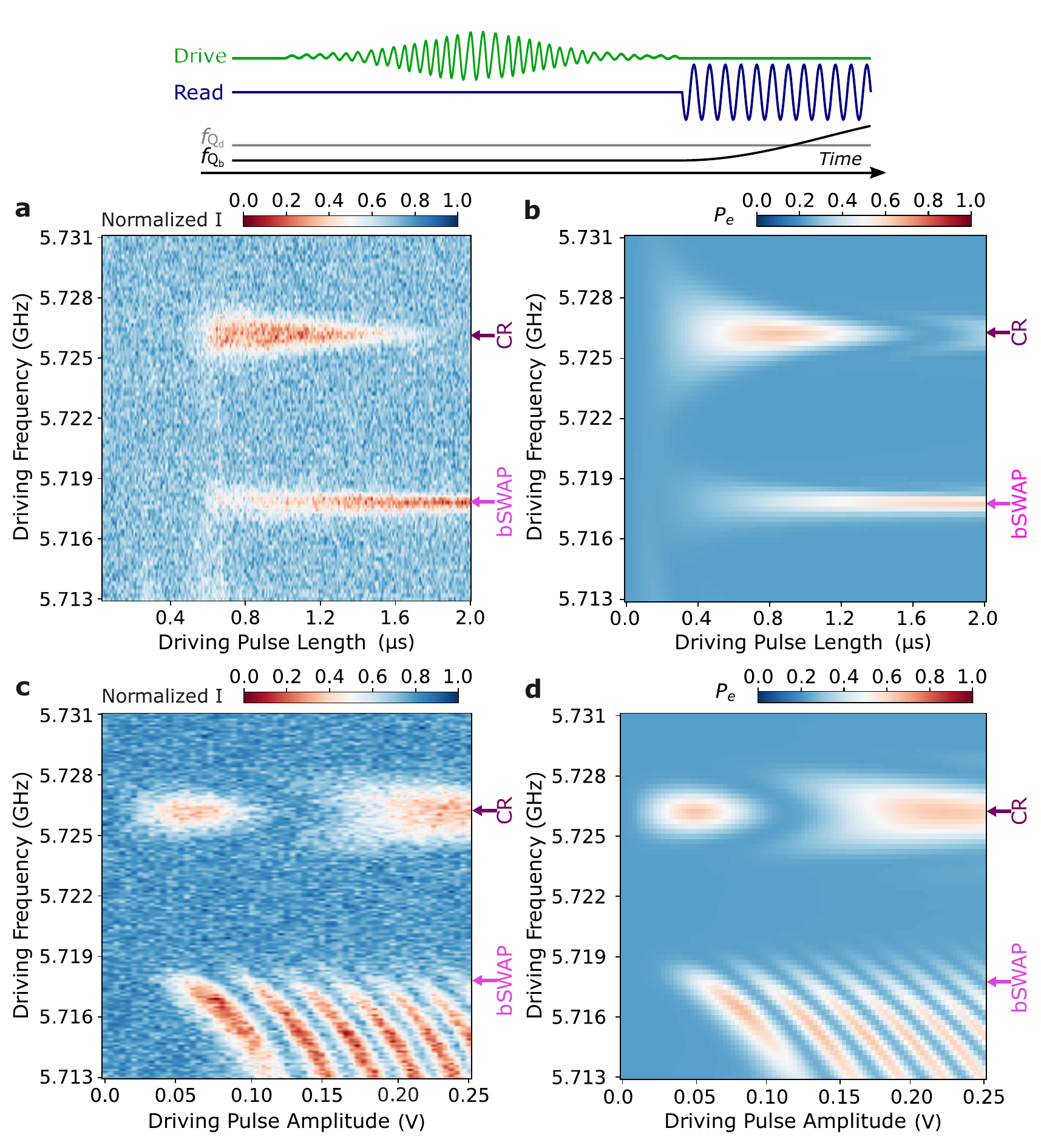}}
	\caption{\textbf{Time-domain all-microwave two-qubit operations on solid neon.} \textbf{a} and \textbf{c,} Drive pulse length-frequency (a) and drive pulse amplitude-frequency (c) Rabi measurements, following the pulse sequences depicted in the inset. In (a), the drive pulse power is fixed at $-70$\,dBm, with varying length from 0 to 2.0\,\textmu s. In (c), the drive pulse length is fixed at 0.8\,µs, with output amplitude varying from 0.0\,V to 0.25\,V, corresponding to zero power to $-57$\,dBm reaching the resonator input coupler. The parameters of the readout pulses are the same as described in the Two-qubit Spectroscopy section. The system is biased with a single DC gate at $\Delta V_\mathrm{tg}=0$. Measurement results show the two oscillations corresponding to cross-resonance (CR, purple arrows) and bSWAP (magenta arrows) two-qubit gates. \textbf{b} and \textbf{d,} Numerically simulated excited state population of $\mathrm{Q_b}$ after the readout pulses, corresponding to the Rabi measurements in (b) and (d). The CR and bSWAP oscillation features are reproduced with experimentally measured parameters from the spectroscopy and decoherence characterization, see Methods for details.} \label{Fig:Rabi}
\end{figure*}

In contrast, the direct interaction between $\mathrm{Q_d}$ and the resonator is not detectable. Meanwhile, the secondary frequency shift of the resonator caused by the dispersive push of $\mathrm{Q_d}$ on $\mathrm{Q_b}$ is also small. Together, these lead to a negligible $\mathrm{Q_d}$-state-dependent resonator response and the absence of visible $\mathrm{Q_d}$ spectrum in Fig.~\ref{Fig:two_qubit_spectrum}a. 
In addition, the spectroscopic symmetry of the two avoided crossings in against $\Delta V_\text{tg}$ indicates that the frequency of the dark qubit $\mathrm{Q_d}$ is much less sensitive to $\Delta V_{\rm tg}$ compared to that of $\mathrm{Q_b}$, remaining nearly constant at 5.726\,GHz over the voltage range scanned in Fig.~\ref{Fig:two_qubit_spectrum}a. These observations reveal that the $\mathrm{Q_d}$'s dipole is nearly orthogonal to both the MW and DC gate fields, mainly defined by the neon-induced structural confinement.

We attribute the observed two-qubit coupling to charge-charge interactions at short distances. Due to the distinct electron-photon coupling strengths of the two qubits individually with the resonator, we can rule out the case of virtual microwave photon exchange between the two qubits through the resonator~\cite{majer2007coupling}. 

Although we cannot readout the dark qubit through the resonator directly, bringing $\mathrm{Q_b}$ and $\mathrm{Q_d}$ on resonance allows their populations to swap, which reveals the state of $\mathrm{Q_d}$ prior to that. This resonance condition is achieved via the ac-Stark shift induced by the readout pulse. Once resonant, the states exchange, and a subsequent measurement of $\mathrm{Q_b}$ reveals the information of the original $\mathrm{Q_d}$ state. A weak pulse provides an insufficient shift, preventing the swap and leaving $\mathrm{Q_d}$ unmeasured, as in the CW two-qubit spectroscopy of Fig.~\ref{Fig:two_qubit_spectrum}a. We simulated the pulsed readout process with varied probe power, as detailed in Methods, to illustrate the state swap dynamics during the 0.7\,µs readout pulse used in our experiments (Extended Data Fig.~4).

With this method, we perform pulsed qubit spectroscopy when biased at $\mathrm{Q_b}$'s SS, as shown in Fig.~\ref{Fig:two_qubit_spectrum}c.
At low drive power of $-73$\,dBm reaching the resonator, we observe two transitions at 5.711\,GHz and 5.726\,GHz, corresponding to the frequencies of $\mathrm{Q_b}$ and $\mathrm{Q_d}$. 
The appearance of $|0_\mathrm{b}0_\mathrm{d}\rangle \rightarrow |0_\mathrm{b}1_\mathrm{d}\rangle$ transition indicates that the long drive pulse pumps $\mathrm{Q_d}$ to a certain population, corresponding to the CR transition of a two-qubit coupled system~\cite{chow2011simple, rigetti2010fully}, as depicted in Fig.~\ref{Fig:two_qubit_picture}c. With increasing drive power, an additional transition peak ($|0_\mathrm{b}0_\mathrm{d}\rangle \rightarrow |1_\mathrm{b}1_\mathrm{d}\rangle$) emerges near the midpoint between the $\mathrm{Q_b}$ and $\mathrm{Q_d}$ frequencies, located at 5.718\,GHz. 
This corresponds to the bSWAP type of two-qubit operation in Fig.~\ref{Fig:two_qubit_picture}c, involving two-photon excitation~\cite{poletto2012entanglement, roth2017analysis}. The fact that $2\omega_{|1_\mathrm{b}1_\mathrm{d}\rangle}\simeq \omega_{|0_\mathrm{b}1_\mathrm{d}\rangle} + \omega_{|1_\mathrm{b}0_\mathrm{d}\rangle}$ suggests the effective longitudinal (ZZ) interaction in the system is negligible~\cite{beysengulov2024coulomb,fors2024comprehensive}.
In summary, through the frequency-domain characterization, we have observed the coherent coupling between two qubits on solid neon, which can support both the CR and the bSWAP types of two-qubit operation.\\
 
\noindent\textbf{Two-qubit gate operations}

\begin{figure*}[htb]
	\centerline{\includegraphics[scale=0.62]{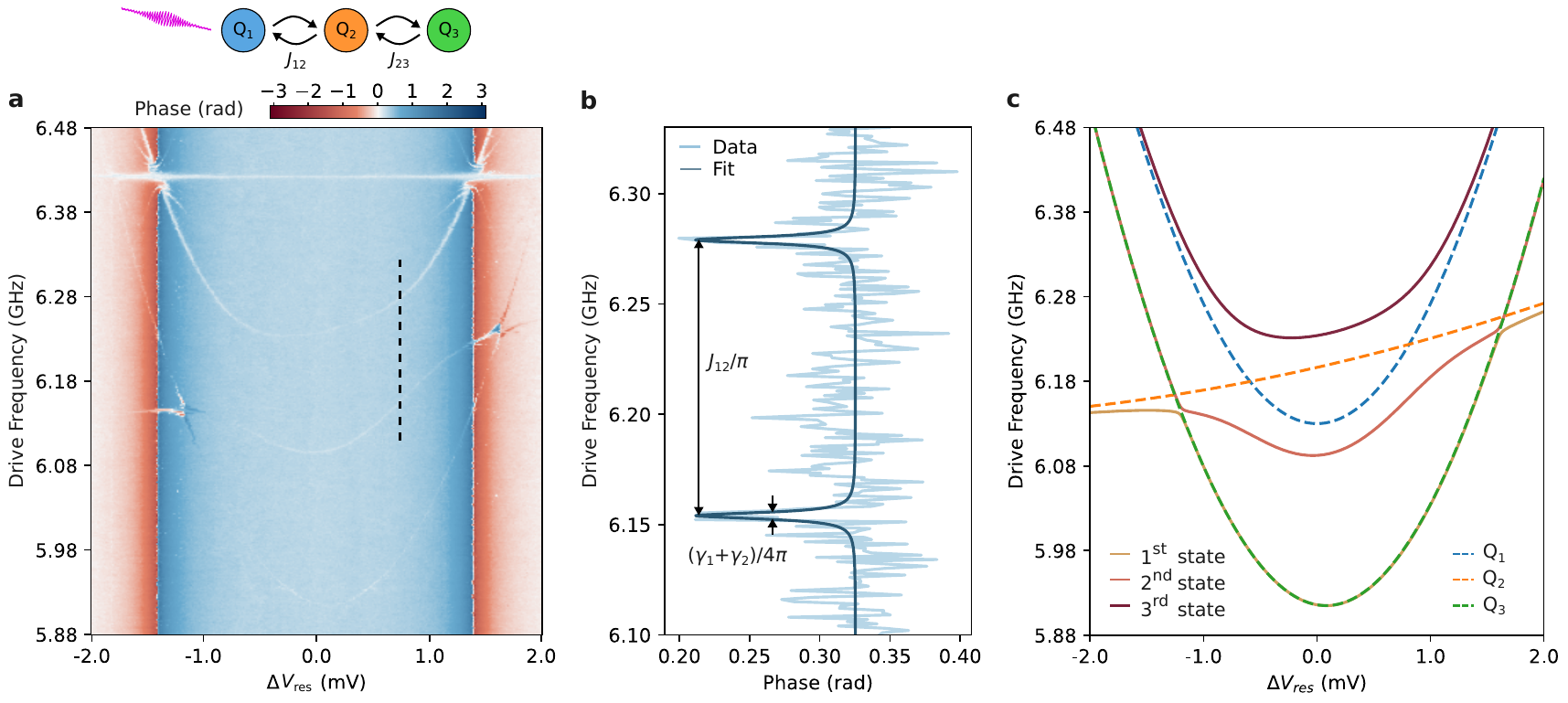}}	\caption{\textbf{Spectroscopic characterization of a three-qubit coupled system on solid neon.} \textbf{a,} Two-tone qubit spectroscopy reveals a three-qubit coupled system, with the inter-qubit coupling strengths $J_\mathrm{12}$ and $J_\mathrm{23}$ shown in the inset. Via $\mathrm{Q}_1$, the system is probed at the resonator frequency while being driven with a second tone at varied frequencies. The dashed line marks the avoided crossing as in (b). \textbf{b,} The line cut shows the large charge-charge interaction strength between $\mathrm{Q}_1$ and $\mathrm{Q}_2$ of $J_\mathrm{12}/2\pi=62.5$\,MHz and hybrid linewidth $(\gamma_1+\gamma_2)/4\pi=3.5$\,MHz. \textbf{c,} Calculated eigenstate energy diagram of the three-qubit coupled system, with bare qubit states shown as dashed lines. See Methods for detailed qubit parameters used in the calculation.} \label{Fig:Spectroscopy}
\end{figure*}

Following the frequency-domain characterization of the coupled system, we perform time-domain two-qubit operations corresponding to the CR and bSWAP gates. A Gaussian-shaped microwave pulse, with variable frequency and duration, excites the system and is followed by a readout pulse at the resonator frequency. This readout pulse also functions as the $\mathrm{Q_b}$ frequency tuning knob with ac-Stark effect and induces population exchange between $\mathrm{Q_d}$ and $\mathrm{Q_b}$. The CR and bSWAP gate operations are manifested as oscillations in the readout signal in Fig.~\ref{Fig:Rabi}a, at 5.726\,GHz and 5.718\,GHz respectively. The simulation of this dynamic process with the system Hamiltonian gives post-readout population of $\mathrm{Q_b}$ with similar patterns, as shown in Fig.~\ref{Fig:Rabi}b. The detailed evolution of the two-qubit system through this drive-readout process is presented in Extended Data Fig.~3.

The simulated results match well with the experimental observations.
For the CR operation (purple arrows), $\mathrm{Q_d}$ was driven by pumping $\mathrm{Q_b}$ at the transition frequency of $\mathrm{Q_d}$, whose excited state population reached maximum when the pulse length was approximately 0.8\,\textmu s. 
Meanwhile, the CR operation barely excites $\mathrm{Q_b}$, as shown in Extended Data Fig.~3. For the bSWAP operation (magenta arrows), the two-photon process drives the two qubits simultaneously, while the oscillation frequency is lower than for the CR operation. When it performs a $\pi$ rotation, which is about 2\,\textmu s in our case, the bSWAP gate is equivalent to an iSWAP gate with appropriate single-qubit operations~\cite{poletto2012entanglement}.

Furthermore, we performed amplitude-dependent Rabi oscillation measurement with a Gaussian-shaped drive pulse of 0.8\,µs duration. Again, we observe oscillating patterns near 5.726\,GHz and 5.718\,GHz, corresponding to the CR (purple arrows) and bSWAP (magenta arrows) operations. 
We attribute the ``waterfall-like" patterns in the frequency range 5.713 through 5.718\,GHz in Fig.~\ref{Fig:Rabi}c to the effects of the Gaussian shape of the drive pulse and the ac-Stark shift of $\mathrm{Q_b}$ under higher drive power, which are reproduced by simulations in Fig.~\ref{Fig:Rabi}d and Extended Data Fig.~3.

These results demonstrate, for the first time, time-domain two-qubit operations in the eNe qubit platform. The neon surface confinement assisted the global all-microwave gate with simplified local DC bias, offering the potential for compact multi-qubit integration on solid neon. Further optimization of the two-qubit gates and readout is crucial to achieving high-fidelity operations.\\

\noindent\textbf{Three-qubit strong coupling}

We further realized a three-qubit system on solid neon. 
The experiment was performed during a separate cooldown using a different resonator from that used in the previous two-qubit system. In this three-qubit system,
$\mathrm{Q}_1$ is directly strong coupled with the resonator, manifested by vacuum Rabi splitting when they are on resonance, while there is no detectable coupling between the other two qubits with the resonator.
The energy diagram of the coupled system was probed via a CW two-tone qubit spectroscopy, while varying the gate voltage on a nearby resonator electrode ($\Delta V_{\rm res}$, see Methods).
The interactions between qubits are revealed by the avoided crossings in the system spectrum, as shown in Fig.~\ref{Fig:Spectroscopy}a. The line cut along the bias point where $\mathrm{Q}_1$ and $\mathrm{Q}_2$ are on resonance, enlarged in Fig.~\ref{Fig:Spectroscopy}b, shows a hybridize-induced avoided crossing with a splitting of $J_\textrm{12}/2\pi$ = 62.5\,MHz. In the same spirit, the coupling strength between $\mathrm{Q}_2$ and $\mathrm{Q}_3$ is approximately $J_\textrm{23}/2\pi$ = 5\,MHz. The difference in inter-qubit coupling strengths arises from variations in the distance and dipole orientation between qubits. Given that the couplings between $\mathrm{Q}_2$ and $\mathrm{Q}_3$ with the resonator are not detectable, likely because of an unfavorable alignment between their dipoles and the microwave field, we attribute the inter-qubit strong coupling to direct charge-charge interactions instead of virtual photon exchange via the resonator.


Figure~\ref{Fig:Spectroscopy}c shows the calculated eigenstates of the system in response to the bias voltage applied on the resonator electrode, aligning with the experimental two-tone spectroscopy. 
The parameters applied in the calculation are listed in Extended Data Tab.~\ref{tab:3-qubit}. 
Due to the large coupling strength between $\mathrm{Q}_1$ and $\mathrm{Q}_2$, the state mixing of the three qubits results in the energy diagram of the coupled system, which deviates from the uncoupled qubit spectrum. 
As a result, the three mixed states are all visible in the two-tone measurement in Fig.~\ref{Fig:Spectroscopy}a. 
These observations indicate the possibilities of simultaneously entangling more than two qubits in the eNe system that may be applied for small- to medium-scale quantum simulations~\cite{wang2023atomic,hensgens2017quantum}. \\

\noindent\textbf{Discussion and outlook}

In this work, we have realized coherent manipulation of multiple electron qubits on solid neon mediated by charge-charge interactions. In the two-qubit system, frequency-domain spectroscopy shows strong inter-qubit coupling, while time-domain measurements demonstrate both cross-resonance and bSWAP two-qubit operations, paving the way for the optimizing high-fidelity two-qubit gates on solid neon. The bright qubit in the system act as the mediator between the dark qubit and the control and readout circuitry, showcasing the engineering potential of neon structures for functioning and scaling eNe qubits with compacted integration and suppressed wiring overheads. Further, the three-qubit system with over 60\,MHz inter-qubit coupling strength shows the potential of building more complex systems with qubits on solid neon. 

Charge-charge interactions between qubits at short distances can enable new possibilities for quantum information processing on solid neon. For the two systems studied in this work, qubits can acquire distinct coupling strengths with superconducting resonators. Leveraging such possibilities, qubits isolated from the resonator can serve as quantum memories for nearby coupled qubits~\cite{wendin2017quantum,xie2024high}, or act as a quantum mediator~\cite{baart2017coherent} in a qubit array. 
Enabling these applications requires further investigation into engineering the electron's structural confinement with solid neon to control qubit properties and their interactions.

Establishing coherent charge interactions between adjacent qubits on solid neon is a crucial step in advancing this emerging solid-state qubit platform. 
However, a multi-qubit system relying only on local interactions will pose challenges to inter-qubit connection efficiency in large-scale integration~\cite{vandersypen2017interfacing,hu2025single}. 
Entanglement between distant qubits through superconducting cavity buses could ease these limits posed by local interactions~\cite{majer2007coupling, dijkema2025cavity, cheung2024photon, bottcher2022parametric}. 
High-impedance resonators made of high-kinetic inductance thin films~\cite{harvey2020chip, harvey2022coherent} and Josephson junction arrays~\cite{stockklauser2017strong, landig2019virtual} have been applied to achieve coherent interactions between distant semiconductor qubits. 
Developing a compatible high-impedance (k$\Omega$ level) resonator~\cite{koolstra2025high, tian2025nbtin}, as well as refined qubit control, would facilitate remote entanglements between distant eNe qubits and hybridize it with other quantum platforms~\cite{xie2024high, pan2025nonlinear}. 
Looking ahead, the integration of short- and long-distance interactions between qubits on solid neon could pave the way for an on-chip quantum network.\\

\setcounter{equation}{0}
\renewcommand{\theequation}{M\arabic{equation}}
\setcounter{figure}{0}
\renewcommand{\thefigure}{\arabic{figure}}
\setcounter{table}{0}
\renewcommand{\thetable}{\arabic{table}}

\renewcommand{\figurename}{Extended Data Fig.}
\renewcommand{\tablename}{Extended Data Tab.}

\noindent\textbf{Methods}

\noindent\textbf{Devices and experiments}

The data presented in Fig.~\ref{Fig:two_qubit_spectrum} and Fig.~\ref{Fig:Rabi} were collected on a high-impedance titanium nitride (TiN) splitting superconducting resonator, the same as the one in ref.~\cite{li2025noise}. A schematic of the electron trap on the resonator is shown in Extended Data Fig.~1a. The resonator supports a differential mode at 5.668\,GHz with a total linewidth of $\kappa/2\pi$~=~0.38\,MHz. Electrodes with on-chip low-pass filters are connected to the resonator pins and trap guards surrounding the electron-trapping region, enabling the application of bias voltages to tune the qubit transition frequency. The device was mounted on a home-made printed circuit board (PCB), sealed in a copper cell with neon filling lines, tungsten filaments, and electrical connectors on top of the cell lid. The cryogenic and room temperature electronics setups are the same as those in ref.~\cite{li2025noise}. The data presented in Fig.~\ref{Fig:Spectroscopy} were collected on a niobium splitting superconducting resonator~\cite{zhou2022single,zhou2024electron}, which supports a differential mode at 6.426\,GHz, with a total linewidth of $\kappa/2\pi$~=~0.46\,MHz. The sample, cryogenic and room temperature electronic setups were the same as those in refs.~\cite{zhou2022single,zhou2024electron}.

The neon growth procedure is the same as that described in ref.~\cite{li2025noise}. In summary, the dilution fridge is warmed up to create a temperature gradient from $\sim$27\,K to $\sim$25\,K between its 4K plate and mixing chamber (MXC) plate. At that moment, neon gas is supplied from a room-temperature gas handling system in the form of puffs and filled onto the device chip in the form of liquid. After a desired amount of neon filling, the fridge heater is turned off to let the fridge cool down to the base temperature.

Electrons are deposited onto the chip at the base temperature by applying pulsed current to the tungsten filaments mounted above the device. During the electron firing process, temperature sensed on the MXC plate can increase up to 100\,mK. The details of the procedure and discussion of the electron deposition process can be found in ref.~\cite{li2025noise}.\\

\noindent\textbf{Modeling multi-qubit coupled systems}

The multi-qubit system Hamiltonian can be written as
\begin{equation}
\hat{H}_{\mathrm{sys}} = \hat{H}_{\mathrm{r}} +\hat{H}_{\mathrm{q}}+\hat{H}_{\mathrm{r\text{-}q}}+\hat{H}_{\mathrm{q\text{-}q}},
\label{Eq:H_s}
\end{equation}
where the resonator Hamiltonian $\hat{H}_\mathrm{r}$, the qubit Hamiltonian $\hat{H}_\mathrm{q}$, the resonator-qubit interaction Hamiltonian $\hat{H}_{\mathrm{r\text{-}q}}$, and the qubit-qubit direct interaction Hamiltonian $\hat{H}_{\mathrm{q\text{-}q}}$ are
\begin{equation}
\hat{H}_{\mathrm{r}}=\hbar\omega_\mathrm{r} \hat{a}^{\dagger}\hat{a},
\end{equation}
\label{Eq:H_r}
\begin{equation}
\hat{H}_{\mathrm{q}} = \sum_{{i}}\frac{1}{2}\hbar\omega_{i}\hat{\sigma}_{{i}}^{\mathrm{z}},
\label{Eq:H_q}
\end{equation}
\begin{equation}
\hat{H}_{\mathrm{r\text{-}q}} = \sum_{{i}}\hbar g_{i} (\hat{a}^{\dagger}\hat{\sigma}_{i}^{-}+\hat{a}\hat{\sigma}_{{i}}^{+}),
\label{Eq:H_rq}
\end{equation}
and
\begin{equation}
\hat{H}_{\mathrm{q\text{-}q}}=\sum_{ij}\hbar J_{ij}(\hat{\sigma}_{i}^{+}\hat{\sigma}_{j}^{-}+\hat{\sigma}_{i}^{-}\hat{\sigma}_{j}^{+})
\label{Eq:H_qq}
\end{equation}
Here, $\omega_\mathrm{r}$ is the resonator frequency, $\omega_{i}$ is the transition frequency of $\mathrm{Q}_i$, $g_{i}$ is the coupling strength between the resonator and $\mathrm{Q}_i$, $\hat{a}$ and $\hat{a}^{\dagger}$ are the annihilation and creation operators of resonator photons, $\hat{\sigma}_i^{z}$, $\hat{\sigma}_i^{-}$, and $\hat{\sigma}_i^{+}$ are the Pauli-$z$, lowering, and raising operators acting on $\mathrm{Q}_i$.
For the dark qubit, we have $g_\mathrm{d}\sim0$. 
On the other hand, the bright qubit has a finite $g_\mathrm{b}$.
$J_{ij}$ represents the transverse inter-qubit coupling strength between $\mathrm{Q}_i$ and $\mathrm{Q}_j$ (ref.~\cite{krantz2019quantum, poletto2012entanglement}). In the Two-qubit Spectroscopy section, we show that the experimentally observed longitudinal (ZZ) coupling strength is small in the system, so we exclude it from the system Hamiltonian model~\cite{fors2024comprehensive, beysengulov2024coulomb}.

The populations of qubits in coupled systems are modeled with their Hamiltonian and simulated using the QuTiP package~\cite{johansson2012qutip}. For the numerical data presented in Fig.~\ref{Fig:Rabi}, we use Eq.~\ref{Eq:H_s} to Eq.~\ref{Eq:H_qq} as the Hamiltonian, with parameters listed in the following sections and Extended Data Tab.~\ref{tab:2-qubit}. The Lindblad master equation is applied to model the effects of relaxation and dephasing:\\
\begin{equation}
\begin{split}
\frac{d\hat{\rho}}{dt}~=~&-\frac{i}{\hbar}[\hat{H},\hat{\rho}] \\
&+\frac{\kappa}{2}(2\hat{a}\hat{\rho}\hat{a}^{\dagger}-\hat{a}^{\dagger}\hat{a}\hat{\rho}-\hat{\rho}\hat{a}^{\dagger}\hat{a}) \\
&+\sum_i\frac{\Gamma_{i}}{2}(2\hat{\sigma}_i^{-}\hat{\rho}\hat{\sigma}_i^{+}-\hat{\sigma}_i^{+}\hat{\sigma}_i^{-}\hat{\rho}-\hat{\rho}\hat{\sigma}_i^{+}\hat{\sigma}_i^{-}) \\
&+\sum_i\frac{\Gamma_{i}^{\varphi}}{2}(2\hat{\sigma}_i^{+}\hat{\sigma}_i^{-}\hat{\rho}\hat{\sigma}_i^{+}\hat{\sigma}_i^{-}-(\hat{\sigma}_i^{+}\hat{\sigma}_i^{-})^2\hat{\rho}-\hat{\rho}(\hat{\sigma}_i^{+}\hat{\sigma}_i^{-})^2), 
\end{split}
\label{Eq:Lindblad}
\end{equation}
where $\hat{H}=\hat{H}_\mathrm{sys}+\hat{H}_\mathrm{dr}$. $\hat{\rho}$ is the system density matrix. $1/\Gamma_{i}$ is the relaxation lifetime $T_1$ of $\mathrm{Q}_i$, and $2/\Gamma_{i}^{\varphi}$ is the pure dephasing time $T_\varphi$ of $\mathrm{Q}_i$, which are adopted from relaxation and Ramsey measurements in Extended Data Fig.~2.

To model the system under drive pulses starting at time $t=0$ with a time-dependent amplitude $A(t)$ and frequency $\omega_\mathrm{dr}$, the driving Hamiltonian can be written as
\begin{equation}
\hat{H}_\mathrm{dr}~=~ A(t)\mathrm{cos}(t\omega_\mathrm{dr})\sum_i\hbar\eta_i(\hat{\sigma}_i^{+}+\hat{\sigma}_i^{-}),
\label{Eq:H_dr}
\end{equation}
where $\eta_i\propto g_i$ is the ratio for direct drive on $\mathrm{Q}_i$ (ref.~\cite{poletto2012entanglement}). In the two-qubit system discussed in the main text $\eta_\mathrm{b}\sim1$ and $\eta_\mathrm{d}\sim0$. The numerical simulation is performed in the rotating frame at the drive frequency.

To model the readout process, we introduce the Hamiltonian of the probing at the resonator ($\hat{H}_\mathrm{p}$) to the system Hamiltonian in the rotating frame as:
\begin{equation}
\hat{H}_\mathrm{p}~=~ \epsilon\hbar(\hat{a}^{\dagger}+\hat{a}),
\label{Eq:H_probe}
\end{equation}
where $\epsilon$ is the probing amplitude.

Details of the experimental characterizations of the system parameters, apart from the ones presented in the main text, are described in the following sections.\\

\noindent\textbf{Characterization of $\mathrm{Q_b}$-resonator coupling}

$\mathrm{Q_b}$ of the two-qubit coupled system presented in the main text is coupled to the resonator with its charge sweet-spot (SS) frequency above the resonator frequency. Due to the interaction, the resonator is dispersively shifted when $\mathrm{Q_b}$ is biased close to its SS, as shown in Extended Data Fig.~1b. The line-cut of the resonator spectrum when ground-state $\mathrm{Q_b}$ is on its SS shows $\chi/2\pi=0.33$\,MHz red shift compared to the bare resonator frequency, as shown in Extended Data Fig.~1c. We estimated the coupling strength between $\mathrm{Q_b}$ and resonator to be $g_\mathrm{b}/2\pi\approx3.76$\,MHz based on this resonator dispersive shift and $\chi\approx g_\mathrm{b}^2/\Delta_\mathrm{b}$ (ref.~\cite{blais2021circuit}), where $\Delta_\mathrm{b}/2\pi = (\omega_\mathrm{b}-\omega_\mathrm{r})/2\pi \approx43$\,MHz.\\

\noindent\textbf{Characterization of qubits decoherence}

We evaluated $\mathrm{Q_b}$ and $\mathrm{Q_d}$'s coherence via relaxation and Ramsey measurements when $\mathrm{Q_b}$ was biased at its SS. $\mathrm{Q_b}$ was directly driven by sending pumps through the resonator at its frequency. In contrast, $\mathrm{Q_d}$ was driven by pumping $\mathrm{Q_b}$ at $\mathrm{Q_d}$'s frequency, corresponding to the cross-resonance two-qubit operation. $\mathrm{Q_d}$ was probed with the pulsed readout method described in the main text and the previous sections. Extended Data Fig.~2a and Extended Data Fig.~2b show the relaxation of $\mathrm{Q_b}$ and $\mathrm{Q_d}$ with $T_1$ of 1.8\,\textmu s and 30.3\,\textmu s, respectively, fitted from the measured in-phase signal. Extended Data Fig.~2c shows the Ramsey fringes of $\mathrm{Q_b}$, with a $T_2^*$ of 2.56\,\textmu s fitted from the in-phase signal. The Ramsey measurement result of $\mathrm{Q_d}$ is more complicated, as we observed multiple frequency components of the measured in-phase signal, as shown in Extended Data Fig.~2d and Extended Data Fig.~2e. Numerical simulation of the Ramsey measurement process with a stable single frequency of $\mathrm{Q_d}$ reveals that the population swap between $\mathrm{Q_b}$ and $\mathrm{Q_d}$ induced by the inter-qubit coupling will not affect the final probed state and measured signal. Therefore, we attribute the beating features in Extended Data Fig.~2d to fluctuations of the $\mathrm{Q_d}$'s frequency during data acquisition, which has also occurred on other qubits measured on the same device, as presented in ref.~\cite{li2025noise}. We fitted the data in Extended Data Fig.~2d with a three-frequency model, which resulted in a $T_2^*$ of 32.99\,\textmu s. The measured relaxation and decoherence rates are then used in the numerical simulation of the system, as shown in Extended Data Tab.~\ref{tab:2-qubit}.\\ 

\noindent\textbf{Summary of simulation parameters}

Extended Data Tab.~\ref{tab:2-qubit} summarizes the simulation parameters to build the Hamiltonian of the two-qubit coupled system and to account for the decoherence. Since the longitudinal interaction is small, as discussed in the main text, ZZ coupling is excluded in the simulation. The resonator has a frequency of 5.668\,GHz with linewidth $\kappa/2\pi=0.38$\,MHz. The interaction strength between the bright qubit and the resonator is $g_\mathrm{b}/2\pi=3.76$\,MHz. In addition, the drive amplitude $A$ in MHz used in the simulation is scaled with $A = 735 \times \mathcal{A}$, where $\mathcal{A}$ is the experimental drive pulse amplitude in $V$. The readout process is simulated with a 0.7\,µs square pulse and probing amplitude $\epsilon$ at 18\,MHz.

Extended Data Fig.~3 shows the simulated qubit excited state population before the readout process in the Rabi measurements, indicated as the arrow time stamp in the inset. The oscillation features at 5.726\,GHz and 5.718\,GHz, corresponding to the cross-resonance (CR) and bSWAP operation, match well with the experimental result in Fig.~\ref{Fig:Rabi} of the main text. It confirms that the observed oscillation features are the result of the two-qubit operations instead of the probing pulse following that. In Extended Data Fig.~3a and Extended Data Fig.~3c, when $\mathrm{Q_d}$ is excited by driving $\mathrm{Q_b}$ at 5.726\,GHz, the $\mathrm{Q_b}$'s population barely changes. After the readout process, the $\mathrm{Q_d}$'s population is swapped onto $\mathrm{Q_b}$, as shown in Fig.~\ref{Fig:Rabi}b and \ref{Fig:Rabi}d of the main text.

The CR and bSWAP oscillations have different frequencies. Assuming a square drive pulse with amplitude $A$, and negligible direct drive on the $\mathrm{Q_d}$, the oscillation frequency induced by CR operation is $\Omega_\text{CR}=AJ/(\Delta_\mathrm{bd}+2J^2/\Delta_\mathrm{bd})$ (ref.~\cite{rigetti2010fully}). 
For the bSWAP operation, the oscillation frequency is approximately $\Omega_\text{bSWAP}=2A^2J/(\Delta_\mathrm{bd}+2J^2/\Delta_\mathrm{bd})^2$ (ref.~\cite{poletto2012entanglement}), where $\Delta_\mathrm{bd}$ is the frequency separation between $\mathrm{Q_b}$ and $\mathrm{Q_d}$. 

To calculate the eigenstates of the three-qubit coupled system, as in the main text Fig.~\ref{Fig:Spectroscopy}, we modeled qubits 1, 2, and 3 with hyperbolic frequency dependency on the gate voltage $\Delta V_{\rm res}$ as: $f_{i} = \sqrt{(\alpha_{i})^2+(\beta_{i}\times(\Delta V_{\rm res}-\delta_{i}))^2}$. In Extended Data Tab.~\ref{tab:3-qubit}, we list the parameters used to generate Fig.~\ref{Fig:Spectroscopy}c in the main text. In addition, we used $J_\textrm{12} /2\pi~=62.5$\,MHz and $J_\textrm{23}/2\pi~=5.0$\,MHz. We also omit longitudinal interactions between qubits in this calculation.\\

\noindent\textbf{Pulsed readout of $\mathrm{Q_d}$}

We demonstrate the pulsed readout process of $\mathrm{Q_d}$ with the following simulation. Initially, the population of $\mathrm{Q_b}$ and $\mathrm{Q_d}$ is set to be 0 and 1, respectively. Then the resonator is populated by sending in a square readout pulse (0.7\,µs) at the resonator frequency with varying amplitude $\epsilon$. The evolution of the coupled system is calculated with the system Hamiltonian Eq.~\ref{Eq:H_s} and the probing Hamiltonian Eq.~\ref{Eq:H_probe}, considering relaxation and dephasing with Eq.~\ref{Eq:Lindblad}.

Extended Data Fig.~4 shows the evolution of the intra-cavity photon number $\bar{n}$, $\mathrm{Q_b}$, and $\mathrm{Q_d}$ excited state population under varying drive amplitude. When the probe power is small (Extended Data Fig.~4a and Extended Data Fig.~4b), the ac-Stark shift of $\mathrm{Q_b}$ caused by the probe pulse is also small, $\mathrm{Q_b}$ and $\mathrm{Q_d}$ weakly exchange population at high frequency approximately $\sqrt{\Delta_\mathrm{bd}^{2}+4J^{2}}$, where $J$ is the inter-qubit coupling strength and $\Delta_\mathrm{bd}$ are the qubits detune. With increased photon number, the blue ac-Stark shift~\cite{blais2021circuit,zhou2024electron} $2\chi\bar{n}$ causes the $\mathrm{Q_b}$ across $\mathrm{Q_d}$ frequency. Given the dispersive coupling strength of $\chi/2\pi\approx0.33$\,MHz, the ac-Stark effect will result in the population swap when $\bar{n}$ approaches 20, as shown in Extended Data Fig.~4c. Further increasing the probing amplitude could populate the resonator faster and make the population swap happen even sooner. These simulation results reveal the readout process we applied to probe $\mathrm{Q_d}$, whose direct interaction strength with the resonator is negligible. We chose the 0.7\,µs readout length to balance signal strength with the fast decaying of $\mathrm{Q_b}$, as illustrated in the qubit decoherence section.\\

\noindent\textbf{Discussion of the ``Waterfall" features in the pulse amplitude-dependent Rabi oscillation}

The ``waterfall-like" pattern measured in pulse amplitude-dependent Rabi oscillation is also replicated in the numerical simulation, by realistically modeling the drive pulse with the same Gaussian envelope used in experiments. We attribute it to the combined effects of the pulse shape and ac-Stark shift of $\mathrm{Q_b}$ under drive at the frequency middle point between $\mathrm{Q_b}$ and $\mathrm{Q_d}$ with high power. Extended Data Fig.~5 plots the simulation when we replace the Gaussian-shaped drive pulse with square pulses. Under such conditions, the ``waterfall-like" pattern vanishes while we could still observe the red shift of the $|0_\mathrm{b}0_\mathrm{d}\rangle \rightarrow |1_\mathrm{b}1_\mathrm{d}\rangle$ transition at higher power, caused by the ac-Stark shift of $\mathrm{Q_b}$. \\

\noindent\textbf{Data availability}

The data that support the findings of this study are available from the corresponding authors upon request. Source data are provided with this paper.\\

\noindent\textbf{Code availability}

The codes used to perform the experiments and to analyze the data in this work are available from the corresponding authors upon request.\\

\acknowledgements

D.J., A.Y., and X.L. acknowledge support from the Air Force Office of Scientific Research (AFOSR) under award no.~FA9550-23-1-0636 for device fabrication and simulation. D.J. acknowledges support from the Department of Energy (DOE) under award no.~DE-SC0025542 for material growth and characterization. D.J. acknowledges support from the National Science Foundation (NSF) under award no.~OSI-2426768 for theoretical modeling. D.J., X.Z., and Y.H. acknowledge support from the Julian Schwinger Foundation for Physics Research for instrument development. X.H. acknowledges support from the U.S. DOE, Office of Science, Advanced Scientific Computing Research (ASCR) program under contract no.~DE-AC02-06CH11357 as part of the InterQnet quantum networking project. Work performed at the Center for Nanoscale Materials, a U.S. Department of Energy Office of Science User Facility, was supported by the U.S. DOE, Office of Basic Energy Sciences, under contract no.~DEAC02-06CH11357. The authors thank Xuedong Hu and David I. Schuster for helpful discussions.\\

\bibliography{eNe_two_qubit}

@article{zhou2022single,
  title={Single electrons on solid neon as a solid-state qubit platform},
  author={Zhou, Xianjing and Koolstra, Gerwin and Zhang, Xufeng and Yang, Ge and Han, Xu and Dizdar, Brennan and Li, Xinhao and Divan, Ralu and Guo, Wei and Murch, Kater W and others},
  journal={Nature},
  volume={605},
  number={7908},
  pages={46--50},
  year={2022},
  publisher={Nature Publishing Group UK London}
}

@article{zhou2024electron,
  title={Electron charge qubit with 0.1 millisecond coherence time},
  author={Zhou, Xianjing and Li, Xinhao and Chen, Qianfan and Koolstra, Gerwin and Yang, Ge and Dizdar, Brennan and Huang, Yizhong and Wang, Christopher S and Han, Xu and Zhang, Xufeng and others},
  journal={Nature Physics},
  volume={20},
  number={1},
  pages={116--122},
  year={2024},
  publisher={Nature Publishing Group UK London}
}

@article{guo2024quantum,
  title={Quantum electronics on quantum liquids and solids},
  author={Guo, Wei and Konstantinov, Denis and Jin, Dafei},
  journal={Progress in Quantum Electronics},
  volume = {99},  
  pages={100552},
  year={2024},
  publisher={Elsevier}
}

@article{jennings2024quantum,
  title={Quantum computing using floating electrons on cryogenic substrates: Potential and challenges},
  author={Jennings, Ash and Zhou, Xianjing and Grytsenko, Ivan and Kawakami, Erika},
  journal={Applied Physics Letters},
  volume={124},
  number={12},
  pages = {120501},
  year={2024},
  publisher={AIP Publishing}
}

@article{platzman1999quantum,
  title={Quantum computing with electrons floating on liquid helium},
  author={Platzman, PM and Dykman, MI},
  journal={Science},
  volume={284},
  number={5422},
  pages={1967--1969},
  year={1999},
  publisher={American Association for the Advancement of Science}
}

@article{burkard2023semiconductor,
  title={Semiconductor spin qubits},
  author={Burkard, Guido and Ladd, Thaddeus D and Pan, Andrew and Nichol, John M and Petta, Jason R},
  journal={Reviews of Modern Physics},
  volume={95},
  number={2},
  pages={025003},
  year={2023},
  publisher={APS}
}

@article{Petta2005coherent,
author = {J. R. Petta  and A. C. Johnson  and J. M. Taylor  and E. A. Laird  and A. Yacoby  and M. D. Lukin  and C. M. Marcus  and M. P. Hanson  and A. C. Gossard },
title = {Coherent Manipulation of Coupled Electron Spins in Semiconductor Quantum Dots},
journal = {Science},
volume = {309},
number = {5744},
pages = {2180-2184},
year = {2005},
doi = {10.1126/science.1116955}
}

@article{zwerver2022qubits,
  title={Qubits made by advanced semiconductor manufacturing},
  author={Zwerver, AMJ and Kr{\"a}henmann, T and Watson, TF and Lampert, Lester and George, Hubert C and Pillarisetty, Ravi and Bojarski, SA and Amin, Payam and Amitonov, SV and Boter, JM and others},
  journal={Nature Electronics},
  volume={5},
  number={3},
  pages={184--190},
  year={2022},
  publisher={Nature Publishing Group UK London}
}

@article{borsoi2024shared,
  title={Shared control of a 16 semiconductor quantum dot crossbar array},
  author={Borsoi, Francesco and Hendrickx, Nico W and John, Valentin and Meyer, Marcel and Motz, Sayr and Van Riggelen, Floor and Sammak, Amir and De Snoo, Sander L and Scappucci, Giordano and Veldhorst, Menno},
  journal={Nature Nanotechnology},
  volume={19},
  number={1},
  pages={21--27},
  year={2024},
  publisher={Nature Publishing Group UK London}
}

@article{vandersypen2017interfacing,
  title={Interfacing spin qubits in quantum dots and donors—hot, dense, and coherent},
  author={Vandersypen, LMK and Bluhm, H and Clarke, JS and Dzurak, AS and Ishihara, R and Morello, A and Reilly, DJ and Schreiber, LR and Veldhorst, M},
  journal={npj Quantum Information},
  volume={3},
  number={1},
  pages={34},
  year={2017},
  publisher={Nature Publishing Group UK London}
}

@article{beysengulov2024coulomb,
  title={Coulomb interaction-driven entanglement of electrons on helium},
  author={Beysengulov, Niyaz R and Sch{\o}yen, {\O}yvind S and Bilek, Stian D and Flaten, Jonas B and Leinonen, Oskar and Hjorth-Jensen, Morten and Pollanen, Johannes and Kristiansen, H{\aa}kon Emil and Stewart, Zachary J and Weidman, Jared D and others},
  journal={PRX Quantum},
  volume={5},
  number={3},
  pages={030324},
  year={2024},
  publisher={APS}
}

@article{kanai2024single,
  title={Single-electron qubits based on quantum ring states on solid neon surface},
  author={Kanai, Toshiaki and Jin, Dafei and Guo, Wei},
  journal={Physical Review Letters},
  volume={132},
  number={25},
  pages={250603},
  year={2024},
  publisher={APS}
}

@article{li2025noise,
  title={Noise-resilient solid host for electron qubits above 100 mK},
  author={Li, Xinhao and Wang, Christopher S and Dizdar, Brennan and Huang, Yizhong and Wen, Yutian and Guo, Wei and Zhang, Xufeng and Han, Xu and Zhou, Xianjing and Jin, Dafei},
  journal={arXiv:2502.01005v2},
  year={2025}
}

@article{zheng2025surface,
  title={Surface-Morphology-Assisted Trapping of Strongly Coupled Electron-on-Neon Charge States},
  author={Zheng, Kaiwen and Song, Xingrui and Murch, Kater W},
  journal={Physical Review Letters},
  volume={135},
  number={8},
  pages={080601},
  year={2025},
  publisher={APS}
}

@article{wang2023atomic,
  title={An atomic-scale multi-qubit platform},
  author={Wang, Yu and Chen, Yi and Bui, Hong T and Wolf, Christoph and Haze, Masahiro and Mier, Cristina and Kim, Jinkyung and Choi, Deung-Jang and Lutz, Christopher P and Bae, Yujeong and others},
  journal={Science},
  volume={382},
  number={6666},
  pages={87--92},
  year={2023},
  publisher={American Association for the Advancement of Science}
}

@article{hensgens2017quantum,
  title={Quantum simulation of a Fermi--Hubbard model using a semiconductor quantum dot array},
  author={Hensgens, Toivo and Fujita, Takafumi and Janssen, Laurens and Li, Xiao and Van Diepen, CJ and Reichl, Christian and Wegscheider, Werner and Das Sarma, Sankar and Vandersypen, Lieven MK},
  journal={Nature},
  volume={548},
  number={7665},
  pages={70--73},
  year={2017},
  publisher={Nature Publishing Group UK London}
}

@article{schuster2010proposal,
  title={Proposal for Manipulating and Detecting Spin and Orbital States of Trapped Electrons on Helium Using Cavity Quantum Electrodynamics},
  author={Schuster, DI and Fragner, A and Dykman, MI and Lyon, SA and Schoelkopf, RJ},
  journal={Physical Review Letters},
  volume={105},
  number={4},
  pages={040503},
  year={2010},
  publisher={APS}
}

@article{blais2021circuit,
  title={Circuit quantum electrodynamics},
  author={Blais, Alexandre and Grimsmo, Arne L and Girvin, Steven M and Wallraff, Andreas},
  journal={Reviews of Modern Physics},
  volume={93},
  number={2},
  pages={025005},
  year={2021},
  publisher={APS}
}

@article{krantz2019quantum,
  title={A quantum engineer's guide to superconducting qubits},
  author={Krantz, Philip and Kjaergaard, Morten and Yan, Fei and Orlando, Terry P and Gustavsson, Simon and Oliver, William D},
  journal={Applied Physics Reviews},
  volume={6},
  number={2},
  pages={021318},
  year={2019},
  publisher={AIP Publishing}
}

@article{roth2017analysis,
  title={Analysis of a parametrically driven exchange-type gate and a two-photon excitation gate between superconducting qubits},
  author={Roth, Marco and Ganzhorn, Marc and Moll, Nikolaj and Filipp, Stefan and Salis, Gian and Schmidt, Sebastian},
  journal={Physical Review A},
  volume={96},
  number={6},
  pages={062323},
  year={2017},
  publisher={APS}
}

@article{rigetti2010fully,
  title={Fully microwave-tunable universal gates in superconducting qubits with linear couplings and fixed transition frequencies},
  author={Rigetti, Chad and Devoret, Michel},
  journal={Physical Review B},
  volume={81},
  number={13},
  pages={134507},
  year={2010},
  publisher={APS}
}

@article{poletto2012entanglement,
  title={Entanglement of Two Superconducting Qubits in a Waveguide Cavity via Monochromatic Two-Photon Excitation},
  author={Poletto, Stefano and Gambetta, Jay M and Merkel, Seth T and Smolin, John A and Chow, Jerry M and C{\'o}rcoles, AD and Keefe, George A and Rothwell, Mary B and Rozen, JR and Abraham, David W and others},
  journal={Physical Review Letters},
  volume={109},
  number={24},
  pages={240505},
  year={2012},
  publisher={APS}
}

@article{koolstra2019coupling,
  title={Coupling a single electron on superfluid helium to a superconducting resonator},
  author={Koolstra, Gerwin and Yang, Ge and Schuster, David I},
  journal={Nature communications},
  volume={10},
  number={1},
  pages={5323},
  year={2019},
  publisher={Nature Publishing Group UK London}
}

@article{chow2011simple,
  title={Simple all-microwave entangling gate for fixed-frequency superconducting qubits},
  author={Chow, Jerry M and C{\'o}rcoles, Antonio D and Gambetta, Jay M and Rigetti, Chad and Johnson, Blake R and Smolin, John A and Rozen, Jim R and Keefe, George A and Rothwell, Mary B and Ketchen, Mark B and others},
  journal={Physical Review Letters},
  volume={107},
  number={8},
  pages={080502},
  year={2011},
  publisher={APS}
}

@article{magesan2020effective,
  title={Effective Hamiltonian models of the cross-resonance gate},
  author={Magesan, Easwar and Gambetta, Jay M},
  journal={Physical Review A},
  volume={101},
  number={5},
  pages={052308},
  year={2020},
  publisher={APS}
}

@article{majer2007coupling,
  title={Coupling superconducting qubits via a cavity bus},
  author={Majer, Johannes and Chow, JM and Gambetta, JM and Koch, Jens and Johnson, BR and Schreier, JA and Frunzio, L and Schuster, DI and Houck, Andrew Addison and Wallraff, Andreas and others},
  journal={Nature},
  volume={449},
  number={7161},
  pages={443--447},
  year={2007},
  publisher={Nature Publishing Group UK London}
}

@article{johansson2012qutip,
  title={QuTiP: An open-source Python framework for the dynamics of open quantum systems},
  author={Johansson, J Robert and Nation, Paul D and Nori, Franco},
  journal={Computer Physics Communications},
  volume={183},
  number={8},
  pages={1760--1772},
  year={2012},
  publisher={Elsevier}
}

@article{wendin2017quantum,
  title={Quantum information processing with superconducting circuits: a review},
  author={Wendin, G{\"o}ran},
  journal={Reports on Progress in Physics},
  volume={80},
  number={10},
  pages={106001},
  year={2017},
  publisher={IOP Publishing}
}

@article{baart2017coherent,
  title={Coherent spin-exchange via a quantum mediator},
  author={Baart, Timothy Alexander and Fujita, Takafumi and Reichl, Christian and Wegscheider, Werner and Vandersypen, Lieven Mark Koenraad},
  journal={Nature Nanotechnology},
  volume={12},
  number={1},
  pages={26--30},
  year={2017},
  publisher={Nature Publishing Group UK London}
}

@article{dijkema2025cavity,
  title={Cavity-mediated iSWAP oscillations between distant spins},
  author={Dijkema, Jurgen and Xue, Xiao and Harvey-Collard, Patrick and Rimbach-Russ, Maximilian and de Snoo, Sander L and Zheng, Guoji and Sammak, Amir and Scappucci, Giordano and Vandersypen, Lieven MK},
  journal={Nature Physics},
  volume={21},
  number={1},
  pages={168--174},
  year={2025},
  publisher={Nature Publishing Group}
}

@article{cheung2024photon,
  title={Photon-mediated long-range coupling of two Andreev pair qubits},
  author={Cheung, LY and Haller, R and Kononov, A and Ciaccia, C and Ungerer, JH and Kanne, T and Nyg{\aa}rd, J and Winkel, P and Reisinger, T and Pop, IM and others},
  journal={Nature Physics},
  volume={20},
  number={11},
  pages={1793--1797},
  year={2024},
  publisher={Nature Publishing Group UK London}
}

@article{harvey2020chip,
  title={On-chip microwave filters for high-impedance resonators with gate-defined quantum dots},
  author={Harvey-Collard, Patrick and Zheng, Guoji and Dijkema, Jurgen and Samkharadze, Nodar and Sammak, Amir and Scappucci, Giordano and Vandersypen, Lieven MK},
  journal={Physical Review Applied},
  volume={14},
  number={3},
  pages={034025},
  year={2020},
  publisher={APS}
}

@article{bottcher2022parametric,
  title={Parametric longitudinal coupling between a high-impedance superconducting resonator and a semiconductor quantum dot singlet-triplet spin qubit},
  author={B{\o}ttcher, CGL and Harvey, SP and Fallahi, Saeed and Gardner, GC and Manfra, MJ and Vool, Uri and Bartlett, SD and Yacoby, Amir},
  journal={Nature Communications},
  volume={13},
  number={1},
  pages={4773},
  year={2022},
  publisher={Nature Publishing Group UK London}
}

@article{harvey2022coherent,
  title={Coherent spin-spin coupling mediated by virtual microwave photons},
  author={Harvey-Collard, Patrick and Dijkema, Jurgen and Zheng, Guoji and Sammak, Amir and Scappucci, Giordano and Vandersypen, Lieven MK},
  journal={Physical Review X},
  volume={12},
  number={2},
  pages={021026},
  year={2022},
  publisher={APS}
}

@article{stockklauser2017strong,
  title={Strong coupling cavity QED with gate-defined double quantum dots enabled by a high impedance resonator},
  author={Stockklauser, Anna and Scarlino, Pasquale and Koski, Jonne V and Gasparinetti, Simone and Andersen, Christian Kraglund and Reichl, Christian and Wegscheider, Werner and Ihn, Thomas and Ensslin, Klaus and Wallraff, Andreas},
  journal={Physical Review X},
  volume={7},
  number={1},
  pages={011030},
  year={2017},
  publisher={APS}
}

@article{landig2019virtual,
  title={Virtual-photon-mediated spin-qubit--transmon coupling},
  author={Landig, Andreas J and Koski, Jonne V and Scarlino, Pasquale and M{\"u}ller, Clemens and Abadillo-Uriel, Jos{\'e} C and Kratochwil, Benedikt and Reichl, Christian and Wegscheider, Werner and Coppersmith, Susan N and Friesen, Mark and others},
  journal={Nature communications},
  volume={10},
  number={1},
  pages={5037},
  year={2019},
  publisher={Nature Publishing Group UK London}
}

@article{xie2024high,
  title={High-fidelity quantum memory with floating electrons coupled to superconducting circuits},
  author={Xie, Ji-kun and Cao, Rong-teng and Ren, Ya-long and Ma, Sheng-li and Zhang, Ren and Li, Fu-li},
  journal={Physical Review A},
  volume={110},
  number={5},
  pages={052607},
  year={2024},
  publisher={APS}
}

@article{pan2025nonlinear,
  title={Nonlinear Tripartite Coupling of Single Electrons on Solid Neon with Magnons in a Hybrid Quantum System},
  author={Pan, Xue-Feng and Li, Peng-Bo},
  journal={arXiv preprint arXiv:2503.08587},
  year={2025}
}

@article{tian2025nbtin,
  title={NbTiN Nanowire Resonators for Spin-Photon Coupling on Solid Neon},
  author={Tian, Y and Grytsenko, I and Jennings, A and Wang, J and Ikegami, H and Zhou, X and Tamate, S and Terai, H and Kutsuma, H and Jin, D and others},
  journal={arXiv preprint arXiv:2505.24303},
  year={2025}
}

@article{koolstra2025high,
  title={High-impedance resonators for strong coupling to an electron on helium},
  author={Koolstra, G and Glen, EO and Beysengulov, NR and Byeon, H and Castoria, KE and Sammon, M and Dizdar, B and Wang, CS and Schuster, DI and Lyon, SA and others},
  journal={Physical Review Applied},
  volume={23},
  number={2},
  pages={024001},
  year={2025},
  publisher={APS}
}

@article{leiderer2025surface,
  title={Surface Electrons on Solid Quantum Substrates: A Brief Review},
  author={Leiderer, Paul},
  journal={Journal of Low Temperature Physics},
  pages={1--20},
  year={2025},
  publisher={Springer}
}

@article{fors2024comprehensive,
  title={Comprehensive explanation of ZZ coupling in superconducting qubits},
  author={Fors, Simon Pettersson and Fern{\'a}ndez-Pend{\'a}s, Jorge and Kockum, Anton Frisk},
  journal={arXiv preprint arXiv:2408.15402},
  year={2024}
}

@article{hu2025single,
  title={Single-Electron Spin Qubits in Silicon for Quantum Computing},
  author={Hu, Guangchong and Huang, Wei Wister and Cai, Ranran and Wang, Lin and Yang, Chih Hwan and Cao, Gang and Xue, Xiao and Huang, Peihao and He, Yu},
  journal={Intelligent Computing},
  volume={4},
  pages={0115},
  year={2025},
  publisher={AAAS}
}

\clearpage

\onecolumngrid

\begin{figure}
  \centering
  \includegraphics[width=1\textwidth]{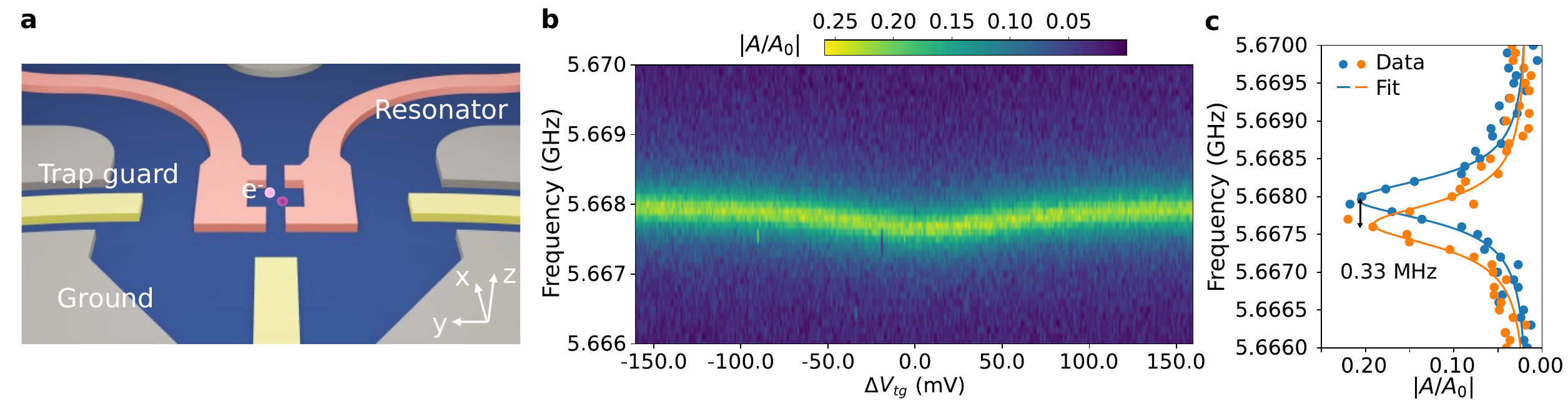}  
  \label{fig:dispersive_shift}
\end{figure}
\twocolumngrid
\noindent\textbf{Extended Data Fig.\,1~\textbar~Resonator device and the dispersive shift caused by $\mathrm{Q_b}$.}
\textbf{a,} Schematic of two closely arranged qubits trapped at the open end of a splitting co-planar waveguide superconducting resonator (pink), surrounded by electrical gates (yellow). The blue and gray structures are the intrinsic silicon substrate and ground, respectively. In the experiment, varying gate voltage is symmetrically applied on the left and right trap guards ($\Delta V_{\text{tg}}$) or on the resonator pins ($\Delta V_{\text{res}}$) to tune the qubits’ transition frequency. \textbf{b,} Resonator spectrum probed with varying bias on the trap guard, showing the dispersive shift when the $\mathrm{Q_b}$ frequency is close to the resonator frequency. \textbf{c,} Transmission line cuts at bare resonator (blue dots and curve) and when $\mathrm{Q_b}$ is biased at its sweet-spot (orange dots and curve.)
\onecolumngrid

\clearpage

\begin{figure}	
	\includegraphics[width=1\textwidth]{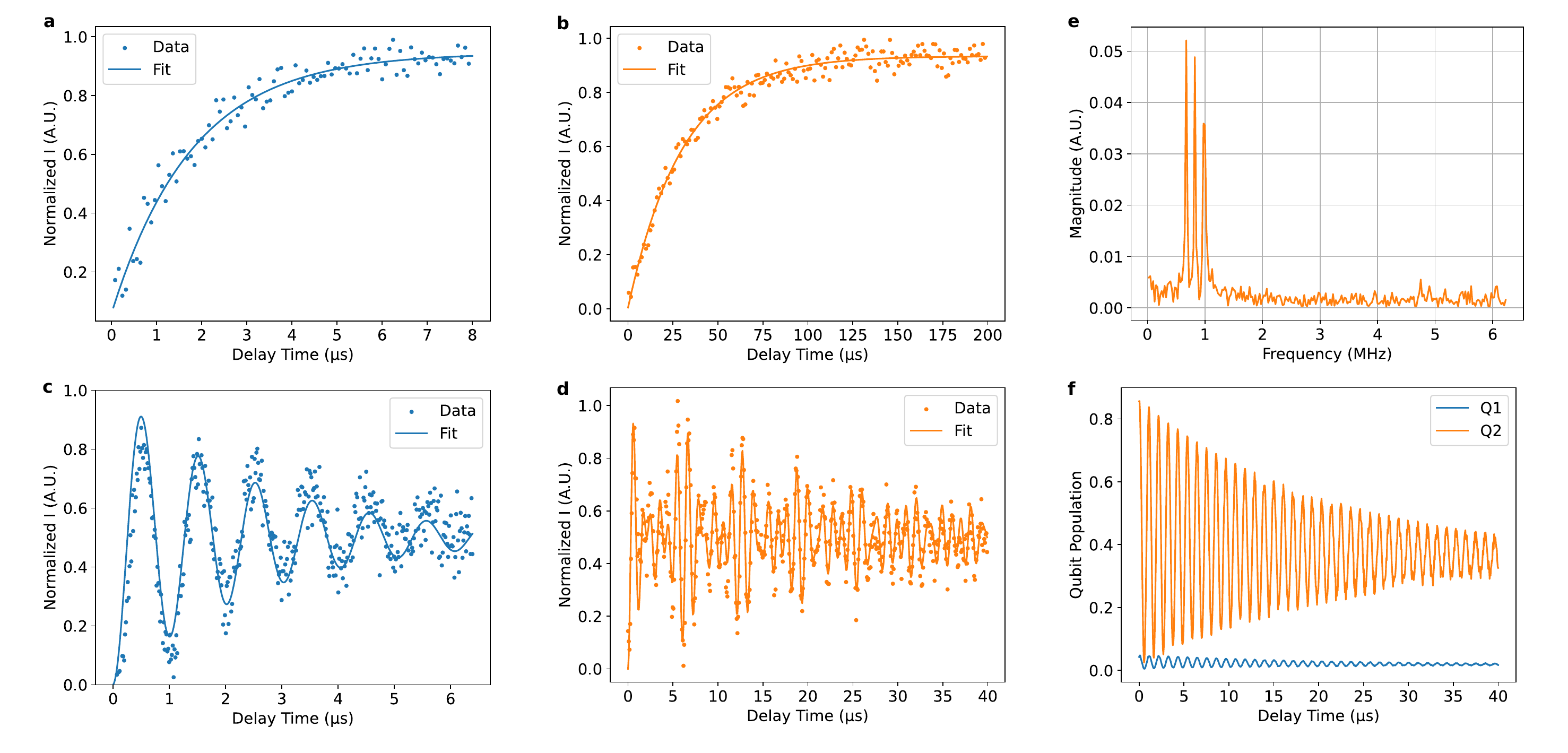}
	\label{Fig:2-qubit decoherence}
\end{figure}
\twocolumngrid
\noindent\textbf{Extended Data Fig.\,2~\textbar~Relaxation and decoherence of $\mathrm{Q_b}$ and $\mathrm{Q_d}$.} \textbf{a} and \textbf{b,} Relaxation measurements of $\mathrm{Q_b}$ (a) and $\mathrm{Q_d}$ (b) with fitted $T_1$ of 1.8\,\textmu s and 30.3\,\textmu s, respectively. \textbf{c} and \textbf{d,} Ramsey measurements of $\mathrm{Q_b}$ (c) and $\mathrm{Q_d}$ (d) with fitted $T_2^*$ of 2.56\,\textmu s and 32.99\,\textmu s, respectively. \textbf{e,} Frequency components of the Ramsey fringes in (d), revealing frequency fluctuations of $\mathrm{Q_d}$ during the measurement. \textbf{f,} Numerical simulation of Ramsey measurement of $\mathrm{Q_d}$, indicating that the beating pattern in (c) is not intrinsically caused by the coupled system.
\onecolumngrid

\clearpage

\begin{figure}
	
	\includegraphics[width=0.6\textwidth]{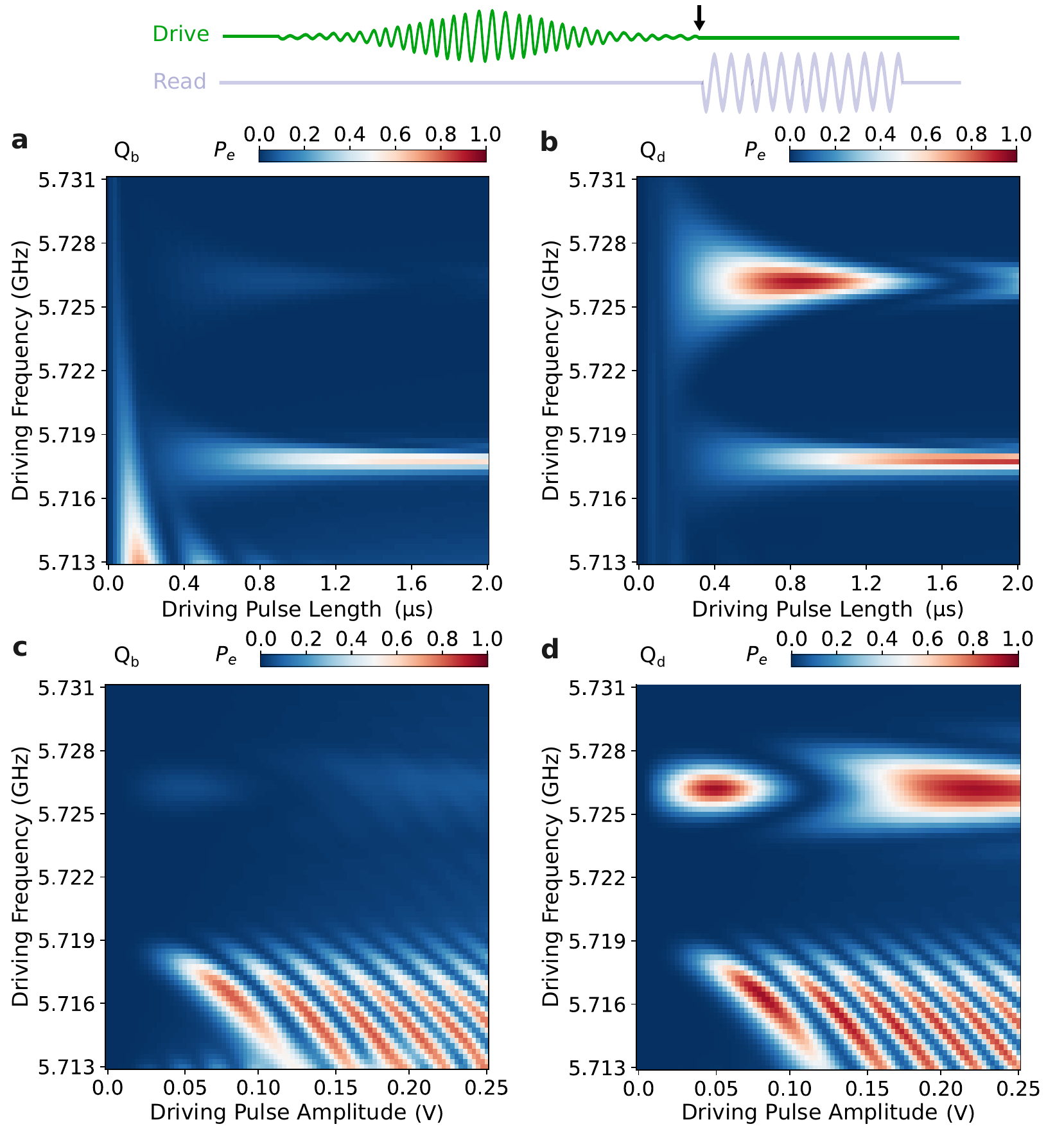}
	\label{Fig: qubit population}
\end{figure}
\twocolumngrid
\noindent\textbf{Extended Data Fig.\,3~\textbar~Simulation results of qubit excited state population before the readout process in Rabi measurements.} \textbf{a} and \textbf{b,} Simulation results of $\mathrm{Q_b}$ (a) and $\mathrm{Q_d}$ (b) population after drive pulse in the pulse length-frequency Rabi measurements. \textbf{c} and \textbf{d,} Simulation results of $\mathrm{Q_b}$ (c) and $\mathrm{Q_d}$ (d) population after drive pulse in the pulse amplitude-frequency Rabi measurements.

\onecolumngrid
\clearpage

\begin{figure}
	\centering
	\includegraphics[width=1\textwidth]{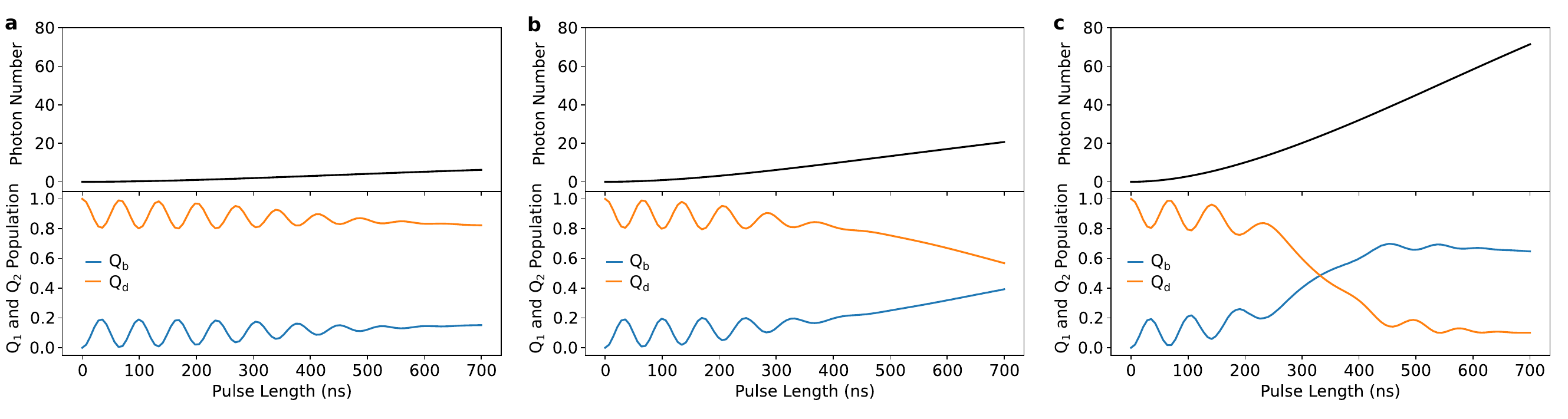}
	\label{Fig: Q2 pulsed readout}
\end{figure}
\twocolumngrid
\noindent\textbf{Extended Data Fig.\,4~\textbar~Simulation of intra-cavity photon number and qubit population evolution with varying probing amplitude.} 0.7\,µs evolution simulated with probing amplitude $\epsilon$ of \textbf{a,} 5.6\,MHz ($-130$\,dBm reaching the resonator), \textbf{b,} 10\,MHz ($-125$\,dBm reaching the resonator) and \textbf{c,} 18\,MHz ($-120$\,dBm reaching the resonator), respectively. Black curves show intra-cavity photon number $\bar{n}$. Blue curves show $\mathrm{Q_b}$ population and orange curves show $\mathrm{Q_d}$ population. 
\onecolumngrid
\clearpage

\begin{figure}
	\centering
	\includegraphics[width=0.6\textwidth]{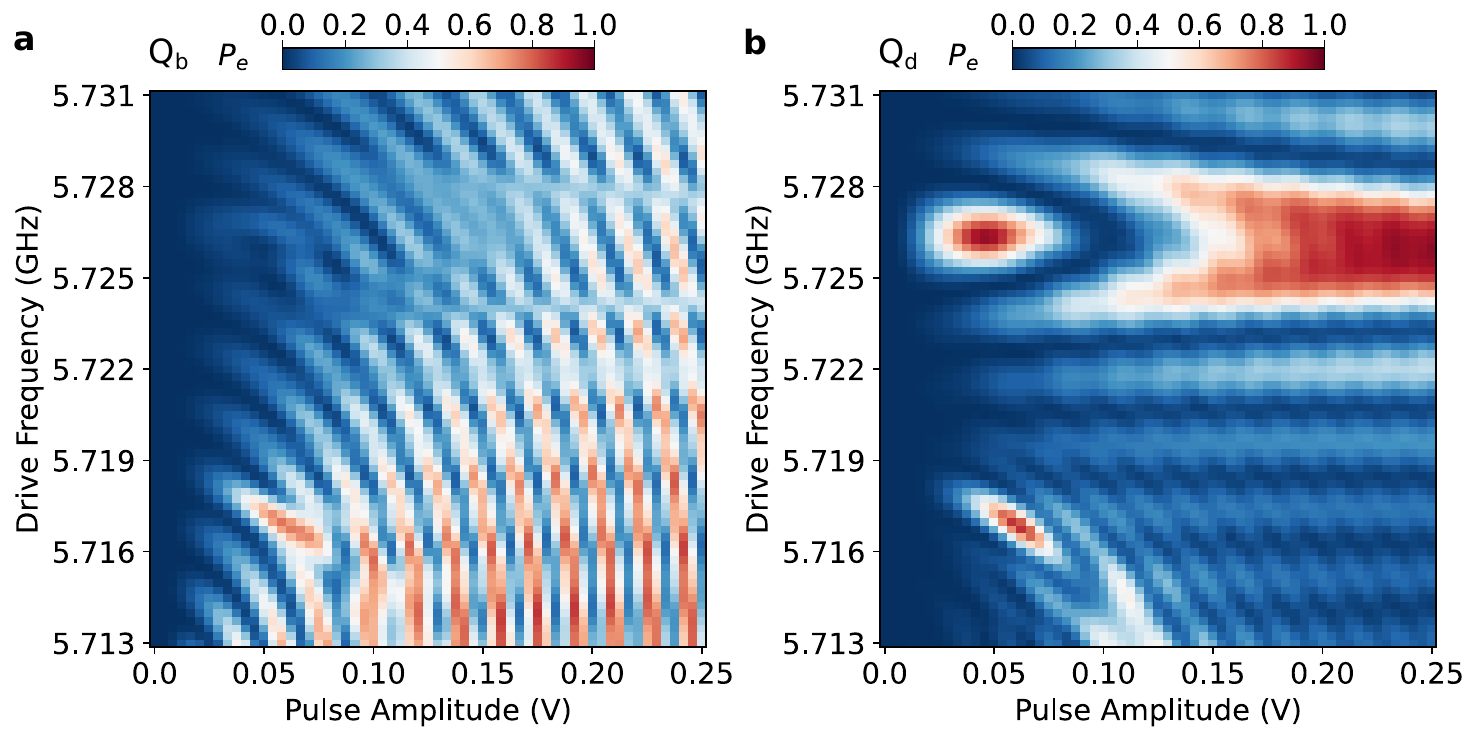}
	\label{Fig:square pulse}
\end{figure}
\twocolumngrid
\noindent\textbf{Extended Data Fig.\,5~\textbar~Numerical simulation of pulse amplitude-dependent Rabi oscillation with step drive pulses.} Simulation results of (a) $\mathrm{Q_b}$, and (b) $\mathrm{Q_d}$ population evolution, in which the ``waterfall-like" pattern observed in experiments vanishes.
\onecolumngrid
\clearpage
\twocolumngrid
\clearpage
\begin{table}
    \caption{\textbar~Summary of the qubit parameters used for simulating the two-qubit coupled system as in Fig.~\ref{Fig:Rabi} and Extended Data Fig.~3.\\}
    \label{tab:2-qubit}
    \centering
    \begin{tabular}{|l|l|l|l|l|l|l|}
    \hline
     & \makecell{$\omega_{i}/2\pi$ \\ (GHz)} & \makecell{$\Gamma_{i}/2\pi$ \\(MHz)} & \makecell{$\Gamma_{i}^{\varphi}/2\pi$ \\ (MHz)} & \makecell{$g_{i}/2\pi$ \\ (MHz)} & \makecell{$J/2\pi$ \\ (MHz)} \\ \hline
    Bright qubit $\mathrm{Q_b}$ & 5.7112 & 0.088 & 0.036 & 3.76 & \smash{\raisebox{-2\height}{\parbox[][2\baselineskip][]{2cm}{3.35}}}  \\ \cline{1-5}
    Dark qubit $\mathrm{Q_d}$ & 5.7255 & 0.0053 & 0.0044 & 0.0 &    \\ \hline
    \end{tabular}
\end{table}

\clearpage

\begin{table}
    \caption{\textbar~Summary of the qubit parameters used for modeling the three-qubit coupled system as in Fig.~\ref{Fig:Spectroscopy}c of the main text.\\}
    \centering
    \begin{tabular}{|l|l|l|l|l|l|l|}
    \hline
     & $\alpha_{i}$ (GHz) & $\beta_{i}$ (GHz/mV) & $\delta_{i}$ (mV)  \\ \hline
    Qubit 1 & 6.130 & 1.325 &0.0  \\ \hline
    Qubit 2 & 6.135 & 0.218 & -4.0   \\ \hline
    Qubit 3 & 5.915 & 1.30 & 0.08 \\  \hline
    \end{tabular}
    \label{tab:3-qubit}
\end{table}

\end{document}